\documentclass[useAMS,usenatbib]{mnras}
\usepackage{times}
\usepackage{graphicx}
\usepackage{epsfig}
\usepackage[T1]{fontenc}
\usepackage{ifpdf}
\usepackage[varg]{txfonts}
\usepackage{hyperref}
\usepackage{deluxetable}

\newcommand{\src}{IGR~J18434}

\def\amin{\ifmmode^{\prime}\else$^{\prime}$\fi}
\def\asec{\ifmmode^{\prime\prime}\else$^{\prime\prime}$\fi}

\def\simgt{\lower.5ex\hbox{$\; \buildrel > \over \sim \;$}}
\def\simlt{\lower.5ex\hbox{$\; \buildrel < \over \sim \;$}}

\newcommand{\goe}{\stackrel{>}{\sim}}

\newcommand\chandra{{\it Chandra}}
\newcommand\xmm{{\it XMM-Newton}}

\newcommand\integral{{\it INTEGRAL}}
\newcommand\INTEGRAL{{\it INTEGRAL}}

\newcommand\swift{{\it Swift\/}}
\newcommand\nustar{{\it NuSTAR}}

\newcommand{\ms}{$M_{\odot}$}

\newcommand{\fluxcgs}{erg~s$^{-1}$~cm$^{-2}$}
\newcommand{\lumcgs}{erg~s$^{-1}$}

\title[X-ray measurement of a high-mass white dwarf and its spin for the intermediate polar IGR J18434-0508]{X-ray measurement of a high-mass white dwarf and its spin for the intermediate polar IGR J18434-0508}

\author[Gerber et al.]{
Julian Gerber$^1$, Jeremy Hare$^{2,9,10}$, John A. Tomsick$^3$, Benjamin M. Coughenour$^{3,11}$, Aarran W. Shaw$^{4}$,
\newauthor
Ma\"{i}ca Clavel$^5$, Francesca Fornasini$^6$, Jules Halpern$^1$, Alyson Joens$^3$, Roman Krivonos$^7$, and Koji Mukai$^8$ \\
$^1$Columbia Astrophysics Laboratory, Columbia University, New York, NY 10027, USA \\ 
$^2$NASA Goddard Space Flight Center, Greenbelt, MD 20771, USA \\
$^3$Space Sciences Laboratory, 7 Gauss Way, University of California, Berkeley, CA 94720-7450, USA \\
$^4$Department of Physics and Astronomy, Butler University, 4600 Sunset Ave, Indianapolis, IN, 46208, USA \\
$^5$Univ. Grenoble Alpes, CNRS, IPAG, 38000 Grenoble, France \\
$^6$Stonehill College, 320 Washington Street, Easton, MA 02357, USA\\
$^7$Space Research Institute, Russian Academy of Sciences, Profsoyuznaya 84/32, 117997 Moscow, Russia\\
$^8$CRESST and X-ray Astrophysics Laboratory, NASA Goddard Space Flight Center, Greenbelt, MD 20771, USA\\
$^9$Center for Research and Exploration in Space Science and Technology, NASA/GSFC, Greenbelt, Maryland 20771, USA \\
$^{10}$The Catholic University of America, 620 Michigan Ave., N.E. Washington, DC 20064, USA\\
$^{11}$Department of Physics, Utah Valley University, 800 W. University Parkway, MS 179, Orem, UT 85048, USA
}
\begin{document}

\def\lsim{\mathrel{\lower .85ex\hbox{\rlap{$\sim$}\raise
.95ex\hbox{$<$} }}}
\def\gsim{\mathrel{\lower .80ex\hbox{\rlap{$\sim$}\raise
.90ex\hbox{$>$} }}}

\pagerange{\pageref{firstpage}--\pageref{lastpage}} \pubyear{2014}

\maketitle

\label{firstpage}

\begin{abstract}

\noindent
IGR J18434$-$0508 is a Galactic Intermediate Polar (IP) type Cataclysmic Variable (CV) previously classified through optical spectroscopy. The source is already known to have a hard \chandra\ spectrum. In this paper, we have used follow-up \xmm\ and \nustar\ observations to measure the white dwarf (WD) mass and spin period. We measure a spin period of $P=304.4\pm0.3$ s based on the combined MOS1, MOS2, and pn light curve. Although this is twice the optical period found previously, we interpret this value to be the true spin period of the WD. The source has an $8\pm2\%$ pulsed fraction in the 0.5--10 keV \xmm\ data and shows strong dips in the soft energy band (0.5--2 keV). The \xmm\ and \nustar\ joint spectrum is consistent with a thermal bremsstrahlung continuum model with an additional partial covering factor, reflection, and Fe-line Gaussian components. Furthermore, we fit the joint spectrum with the post-shock region ``ipolar" model which indicates a high WD mass $\goe$1.36 \ms, approaching the Chandrasekhar limit. 
\end{abstract}

\section{Introduction}
\label{Intro}
Since its launch in 2002, the International Gamma-Ray Astrophysics Laboratory (\INTEGRAL) has resulted in a number of hard X-ray catalogs surveying the full sky \citep{Bird2010, Bird2016} and the Galactic Plane \citep{INTEGRAL14year} from $17-60$ keV; see \cite{Krivonos2021} for a review on \integral's survey. The most recent such survey, \cite{INTEGRAL17year}, is a catalogue of 929 ``IGR" sources detected across 17 years of \integral\ observations. While most of these sources have been identified as either Active Galactic Nuclei (AGN), X-Ray Binaries, or Cataclysmic Variables (CVs), there are over 100 sources yet to be classified. The strategy to classify these sources involves first finding a soft X-ray counterpart to the IGR source, which helps to localize the source better than the few arcminutes accuracy of \integral. With better positional accuracy, follow-up observations may be taken to classify the source. For example, this strategy was employed to identify IGR J18007$-$4146, IGR J14091$-$6108, IGR J15038$-$6021, and IGR J17528$-$2022 as Intermediate Polars by \cite{Coughenour2022}, \cite{Tomsick2016}, \cite{Tomsick2023}, and \cite{Hare2021} respectively.

CVs are binary systems containing a white dwarf (WD) accreting from a main sequence companion star. Intermediate Polars (IPs) are a class of CVs (magnetic CVs, or mCVs, in particular) in which the magnetic field of the WD truncates the accretion disc surrounding it. This causes the material to funnel along the magnetic field lines of the WD towards its magnetic poles. Polars are mCVs which have even stronger magnetic fields, such that no accretion disc forms around the WD at all. CVs, which accrete via Roche-lobe overflow, are copious emitters of X-rays which make them good targets to study with hard X-ray observatories such as \INTEGRAL\ \citep{Mukai2017, LUTOVINOV2020}.

IGR J18434$-$0508 (hereafter, \src) was detected in the 14-year \INTEGRAL\ hard X-ray survey of the Galaxy at R.A. = $280.855^{\circ}$, DEC = $-5.138^{\circ}$ \citep{Krivonos2017}. A counterpart search was conducted by \cite{Krivonos2017} within the source's 90\% confidence error circle. Both hard and soft X-ray counterparts were found using \swift/BAT and \swift/XRT. These counterparts are 4PBC~J1842.8$-$0506 and Swift~J184311.0$-$050539, respectively.

In a follow-up to the 14-year \INTEGRAL\ survey \citep{Krivonos2017}, \cite{Tomsick2021} identified unique \chandra\ counterparts to several of the new IGR sources with a high degree of confidence, and one of which was \src. The probability of a match between the \integral\ and \chandra\ sources was calculated as a function of number of counts between 2 and 10 keV and angular distance between the sources. The \chandra\ counterpart to \src\ was determined to be CXOU~J184311.4$-$050545 with a more accurate position of R.A. = $18^{\rm h}43^{\rm m}11.43^{\rm s}$, DEC = $-05^{\circ}05^{\prime}45.2^{\prime\prime}$. 
The sub-arcsecond \chandra\ positional accuracy allowed \cite{Tomsick2021} to also find a {\it Gaia} and UKIDSS counterpart to \src\ using the Gaia EDR3 Catalog and the UKIRT Infrared Deep Sky Survey in the VizieR database \citep{GaiaEDR32021, Lucas2008}. The Gaia counterpart to \src\ has a distance of $3.0_{-0.9}^{+1.2}$ kpc \citep{BailerJones2021} which corresponds to a 2--10 keV luminosity of $4.2_{-2.6}^{+3.4}\times10^{33}$ \lumcgs \citep{Tomsick2021}. The hardness of the \chandra\ spectrum that was extracted (with a power-law photon index of $\Gamma=0.7\pm0.3$) indicated that \src\ was either a mCV or a high mass X-ray binary (HMXB). Using the $J$, $H$, and $K$ magnitudes of the UKIDSS counterpart (UKIDSS~J184311.43-050545.6) as well as the stellar color tables in \cite{Pecaut&Mamajek2013}, it was found that \src\ contained a late type donor star, eliminating the possibility of an HMXB. It was therefore concluded by \cite{Tomsick2021} that \src\ was a strong mCV candidate. In the most recent study discussing \src, \cite{Halpern2022} conducted optical spectroscopy and confirmed that \src\ is a CV based on its broad H$\alpha$ emission line. In that same work, time-series photometry found a period of $152.49\pm0.02$~s.

The primary new information we report in this paper is timing and spectral analysis of \src\ using \xmm\ and \nustar, which we use to determine the mass of the white dwarf in the CV and its true spin period. In Section \ref{Obs}, we describe the \xmm\ and \nustar\ observations taken of \src\ as well as the data reduction.  In Section \ref{timing}, we discuss the X-ray timing analysis of \src\ which found a period consistent with twice the optical period. In Section \ref{spectral}, we discuss the spectral analysis results which established the high mass nature of \src. Finally, in Section \ref{discussion}, we discuss the source identification process for \src\ and conclude that it is a high mass Intermediate Polar. 
\section{Observations and Data Reduction}
\label{Obs}
Observations of \src\ were conducted using \nustar\ and \xmm. The observations were taken simultaneously, beginning on 2022 March 12. Details on these observations are listed in Table \ref{obstab}.

\subsection{\xmm}

We reduced the EPIC/pn and EPIC/MOS data using the \xmm\ Science Analysis Software (SAS v20.0). We first ran the SAS {\tt emchain} and {\tt epchain} scripts to generate the processed event lists for MOS and pn. For pn, we then filtered the event list using {\tt evselect} based on the expression ``(PATTERN $<$= 4)\&\&(PI in [200:15000])\&\&(FLAG == 0).'' We also filtered MOS1/2 with the expression ``(PATTERN $<$= 12)\&\&(PI in [200:12000])\&\&\#XMMEA\_EM.'' We further filtered the MOS event files for soft proton (SP) flares using the {\tt mos-filter} script. The pn observation was conducted in Small Window mode and was thus not supported by {\tt pn-filter}. We therefore extracted a pn light curve using {\tt evselect} in order to manually filter for SP contamination. After finding no such flares in the pn light curve, we simply filtered the pn event list by the original good time interval (GTI) file produced by {\tt epchain}. Additionally, the event arrival times for the pn, MOS1, and MOS2 detectors were corrected to the solar system barycenter for timing analysis.

We then extracted three source spectra from a circular region of radius $30^{\prime\prime}$ centered at the source position for each instrument. We did the same for annular background regions with outer radii of $50^{\prime\prime}$ and inner radii of $32.5^{\prime\prime}$. We then used {\tt backscale} to account for the size of our extraction regions. Furthermore, we used {\tt rmfgen} and {\tt arfgen} to create the response matrices required for spectral fitting. Finally, we grouped the source spectra to contain a minimum signal-to-noise ratio of $5\sigma$ using {\tt ftgrouppha}.

\subsection{\nustar}

The Level 1 science event files for ObsID 30760002002 were reduced using NUSTARDAS v2.1.1 and CALDB 20220215. The analysis was done using HEASoft version 6.30.1. Source and background regions were defined using the FPMA/B Level 2 event files. The source regions were both circular with radii of $49.2^{\prime\prime}$ (equal to 20 pixels). The backgrounds were rectangular regions offset from the source but on the same detector chip. Using these regions, we ran {\tt nuproducts} for FPMA/B to create source and background spectra as well as response matrices. Like the \xmm\ spectra, the \nustar\ source spectra were binned using {\tt ftgrouppha} to have at least a $5\sigma$ signal-to-noise ratio. We also corrected the {\it \nustar\/} event arrival times to the solar system barycenter. 

\section{X-ray Timing}
\label{timing}
\begin{figure*}
\begin{center}
    \includegraphics[width=15cm]{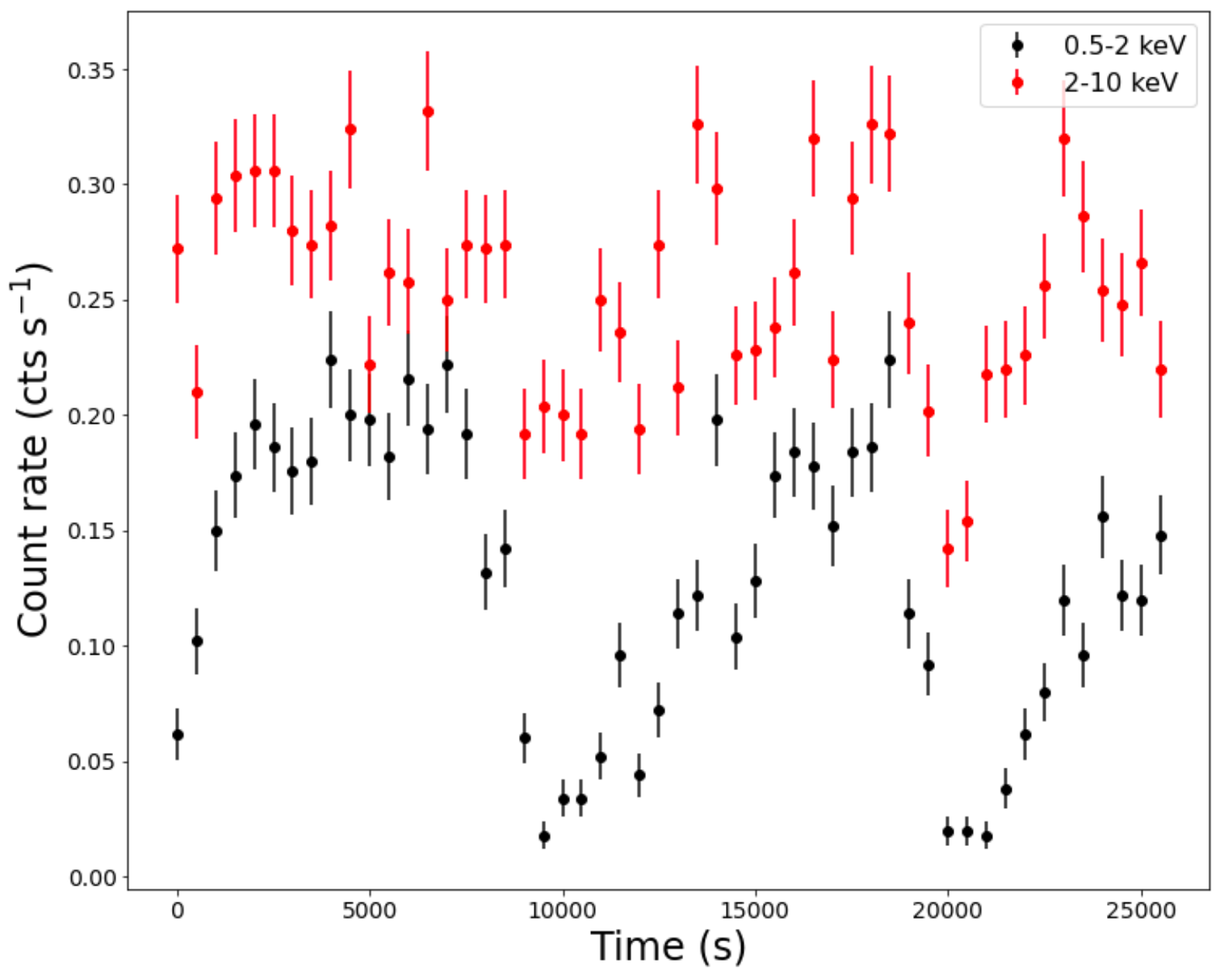}
    \caption{\xmm\ EPIC pn 0.5-2 (black) and 2-10 (red) keV light curves with a 500 s binning. Variability is seen in both light curves, but is much more visible in the soft light curve.}
    \label{pn_05-2_2_10LC}
\end{center}
\end{figure*}

We used the barycenter corrected times to create light curves of the source with \xmm\  and \nustar\ (extracted from $r=30''$ and $r=50''$ circular regions, respectively).
Notably, two strong dips separated by about 10 ks are observed. The source also appears to be rising out of a dip at the start of the observation. To further explore these dips, we extracted the pn light curves in soft (0.5-2 keV) and hard (2-10 keV) energy ranges (see Figure \ref{pn_05-2_2_10LC}).
The energy resolved light curves show that the dips are much more prominent at soft energies compared to the hard  energies. To search for a potential orbital period in the X-ray light curve, we calculated the Lomb-Scargle periodogram using the full, soft, and hard 500 s binned X-ray light curves. A strong peak is observed around a period of about 3.1 hours (however, see Section 5 for further discussion).
Assuming this is the correct peak in the power spectrum, we estimate an uncertainty of 0.8 hr on the period from the full width at half-max of the peak in the periodogram.

We also constructed \nustar\ light curves with a 1 ks binning in the full 3-79 keV energy band and the 3-10 keV energy band, since the latter overlaps with the \xmm\ energy range. However, no variability is observed in either light curve. We do see evidence of variability in the 3--10 keV \xmm\ light curve, suggesting that the lack of variability in the 3--10 keV \nustar\ light curve is due to the source having a lower signal to noise ratio in \nustar. It is also possible that the variability observed in the \xmm\ light curve is primarily due to lower energy photons. In this case, the lack of variability in the \nustar\ light curve would be due to the decreased effective area of \nustar\ at lower energies \citep{Harrison2013}.

\begin{figure*}
\begin{center}
\includegraphics[scale=.5]{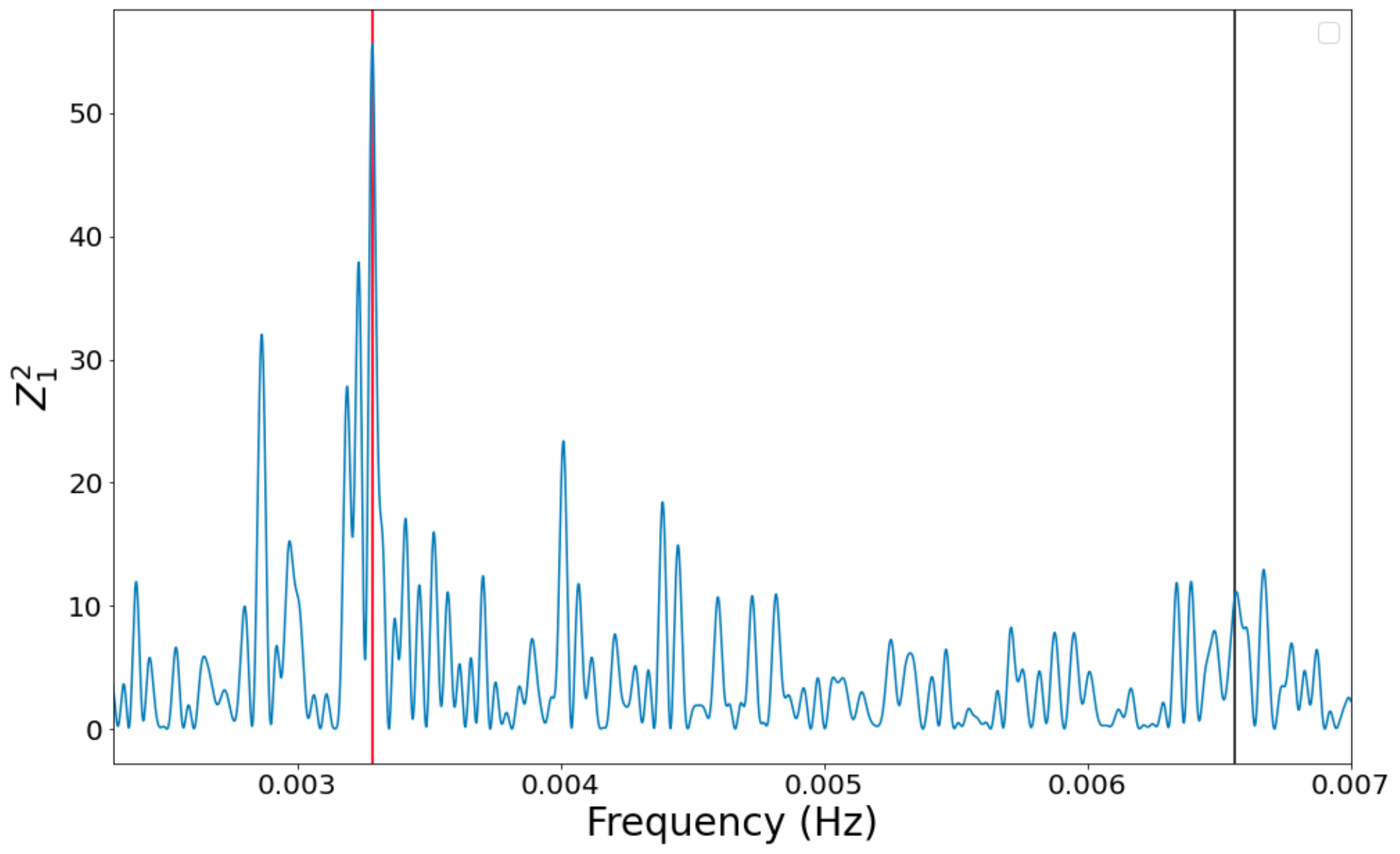}
\caption{$Z_1^{2}$ versus frequency for the \xmm\ combined MOS1, MOS2, and pn, 0.5-10 keV event list. The red line shows the largest peak at $\sim$304 s, while the black peak shows the location of the 152.49 s period observed in the optical power spectrum by \protect\cite{Halpern2022}.}
\label{z1sq_xmm}
\end{center}
\end{figure*}

A periodicity of 152.49$\pm0.02$ s was observed in the optical light curve of IGR J18434 \citep{Halpern2022}. We use the $Z_1^{2}$ test \citep{Buccheri1983} to search for pulsations at this period using the \xmm\ and \nustar\ event lists. First, we searched for the 152.49 s period using the combined (MOS1+MOS2+pn) event list. The largest peak found in the immediate vicinity (i.e., between $\nu=6.47\times10^{-3}$ Hz to $\nu=6.64\times10^{-3}$ Hz or within $\pm100\sigma$) of the optical period is located at 0.00657 Hz (152.3 s) but has a low $Z_1^{2}=11.1$, which is statistically insignificant after accounting for the number of trials.  Expanding the search to a larger frequency range ($\nu=0.0005-0.01$ Hz) uncovers a much stronger period ($Z_1^{2}=55.6$) at a frequency of $\nu=0.003285\pm0.000003$ Hz or $P=304.4\pm0.3$ s, having a False Alarm Probability (FAP) of $\sim$10$^{-8}$. The results of the \xmm\ $Z_1^2$ test are shown in Figure \ref{z1sq_xmm}. The uncertainties are estimated by calculating at which frequency $Z^{2}_{\rm 1,max}$ falls to $Z^{2}_{\rm 1,max}-1$. This period is about twice the observed optical period.  

\begin{figure*}
\begin{center}
    \includegraphics[scale=.5]{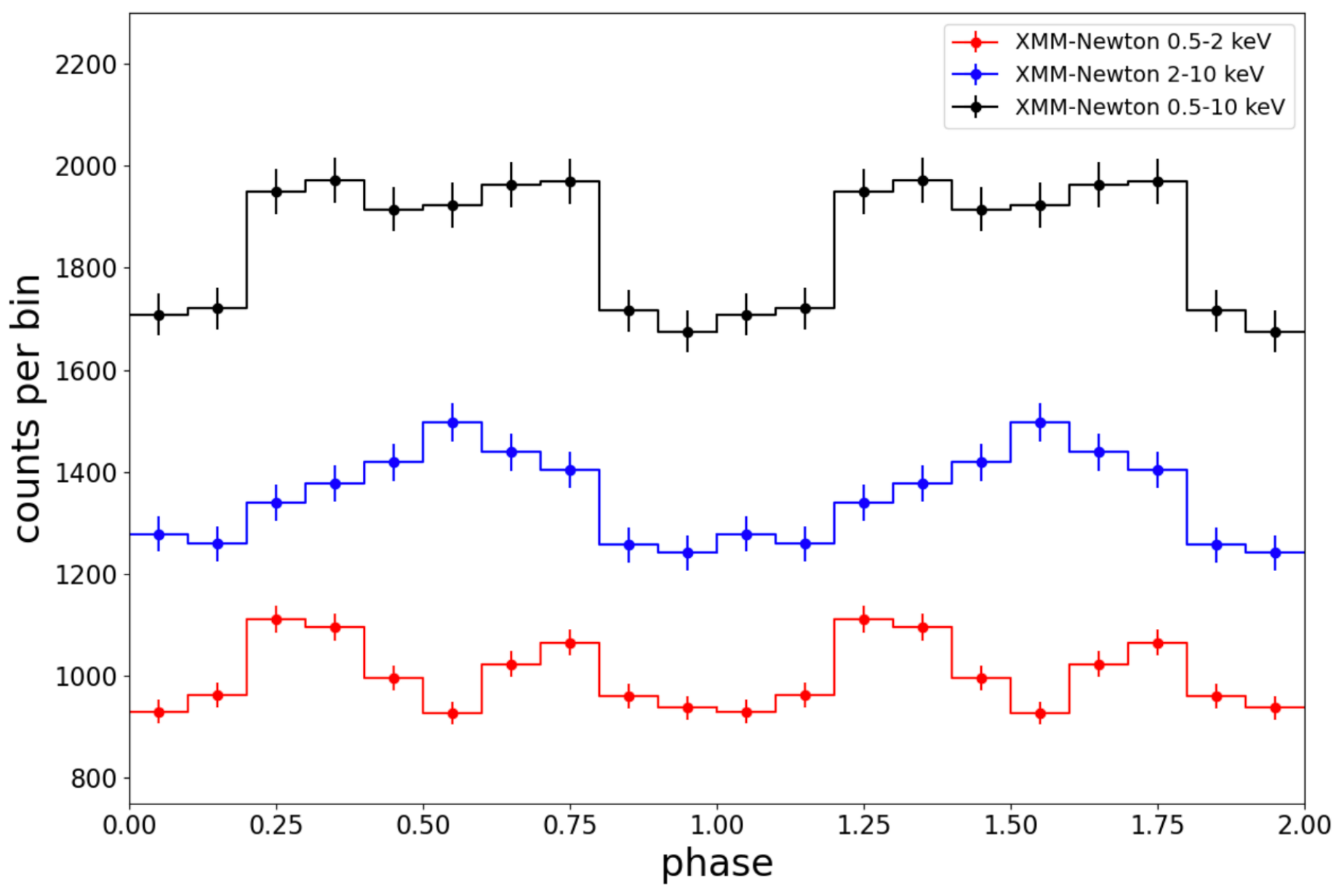}
    \caption{\xmm\ combined MOS1, MOS2, and pn, 0.5-10 keV pulse profile folded on the pulse period of 304.4 s. Two phases are shown for clarity. The soft and hard band pulse profiles have an arbitrary shift applied for easier comparison between the pulse profiles.}
    \label{XMM_pulse_prof}
\end{center}
\end{figure*}

\begin{figure*}
\begin{center}
    \includegraphics[scale=.5]{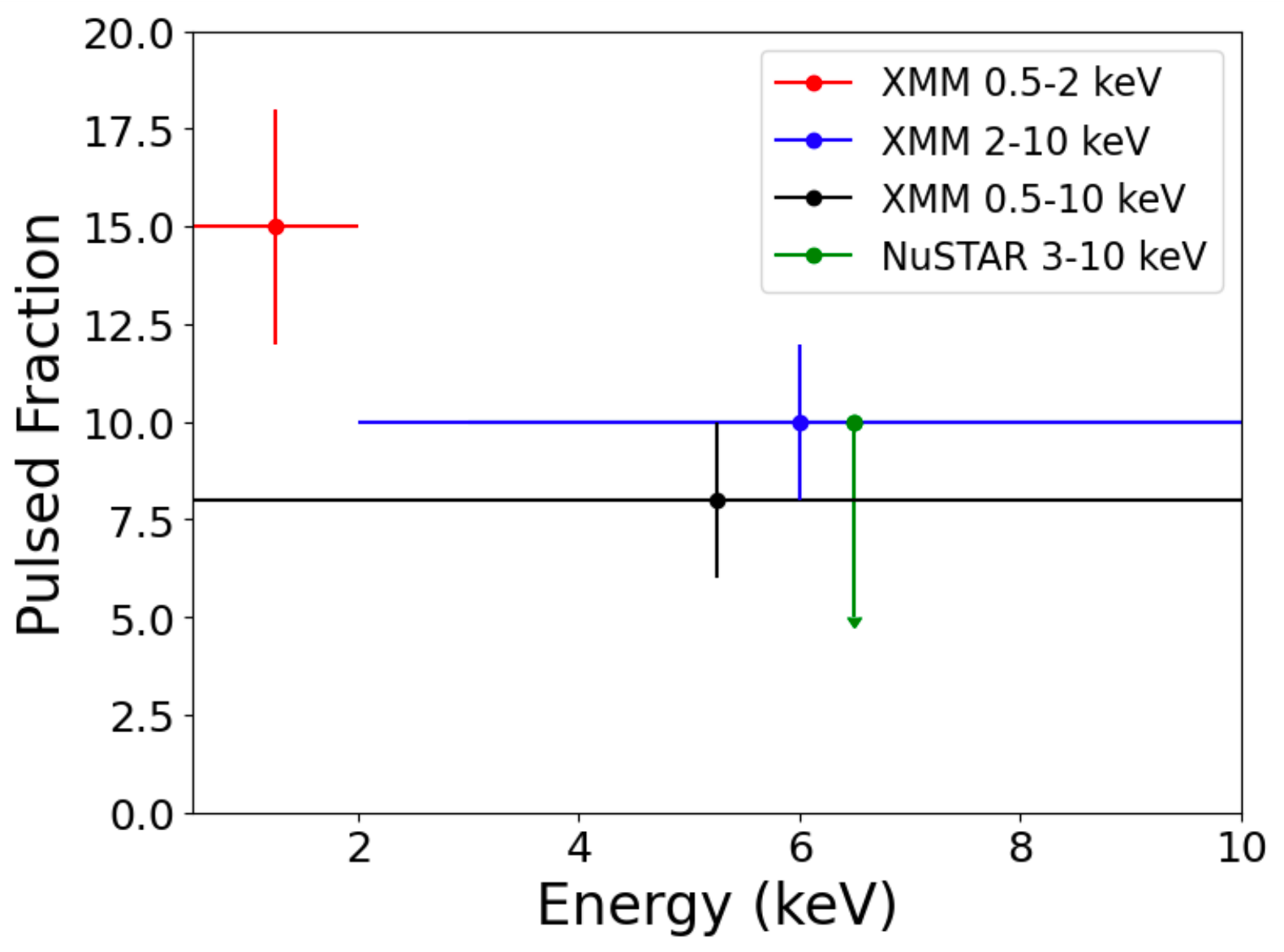}
    \caption{The measured pulsed fraction in several different energy bands. A 3$\sigma$ upper-limit is provided for the NuSTAR non-detection. }
    \label{pulsed_fraction}
\end{center}
\end{figure*}

The 0.5-10 keV pulse profile shows only one peak per period with a relatively flat top (see Figure \ref{XMM_pulse_prof}). We calculate the pulsed fraction, defined as $(C_{\rm max}-C_{\rm min})/(C_{\rm max}+C_{\rm min})$ where $C_{\rm max}$ and $C_{\rm min}$ are the maximum and minimum number of counts in the folded pulse profile, and find a value of $8\pm2\%$ in the 0.5-10 keV band. We also divided the \xmm\ pulse profile into soft (0.5-2 keV) and hard (2-10 keV) energy bands (see Figure \ref{XMM_pulse_prof}). Interestingly, the pulse profiles are markedly different in these energy ranges, with the soft profile showing two peaks per pulse period, while the hard profile shows only one peak per 
period. Additionally, the peak of the hard band pulse profile falls in the minimum between the the two peaks seen in the soft band pulse profile. The soft and hard bands have pulsed fractions of $15\pm3\%$ and $10\pm2\%$, respectively. The observed pulsed fractions as a function of energy are plotted in Figure \ref{pulsed_fraction}.

We also searched for the 304.4 s period in the 3-10 keV \nustar\ data. However, we note that at low frequencies, the gaps in the data caused by Earth occultations due to \nustar's\ low-earth orbit lead to many spurious peaks in the $Z_1^{2}$ periodogram related to the harmonics of \nustar's\ orbital period. Unfortunately one of these harmonics falls very close to the pulse period, making it difficult to determine the significance of the spin period. As a secondary check, we also used the 10 s binned \nustar\ light curve and Stingray \citep{2019ApJ...881...39H,Huppenkothen2019} to calculate the average power spectra over continuous good time intervals longer than 3 ks, thus removing the power in the harmonics caused by \nustar's orbit. No statistically significant peak is detected at or near the expected pulse frequency. We place a 3$\sigma$ upper-limit on the observed pulsed fraction of $10\%$ in the 3-10 keV energy band.

\section{X-ray Spectrum}
\label{spectral}
For spectral analysis we jointly fit the \xmm\ and \nustar\ data in the 0.5--79 keV range (0.5--12 keV for \xmm\ and 3--79 keV for \nustar) using the XSPEC spectral modeling package \citep{Arnaud1996}. In order to account for differences in normalization across instruments, we fit each spectral model using a multiplicative constant. The constant for MOS1 was frozen to 1 while the others were allowed to fit freely. Furthermore, {\tt{tbabs}} was included in all fits in order to account for ISM absorption. The abundances for the {\tt{tbabs}} component may be found in \cite{Wilms2000}. For all parameters listed, the errors shown are the $1\sigma$ confidence intervals.

We initially fit both a power-law and a thermal bremsstrahlung model to the \src\ data. However, in both cases, the resulting $\chi_{\rm red}^2$ is unacceptably high at 2.4 and 2.5, respectively, for 820 degrees of freedom (dof). In order to improve the bremsstrahlung fit, we included the {\tt{reflect}} model described in \cite{Magdziarz1995}. This component modifies the model to account for reflection off of the surface of the WD. While the bremsstrahlung fit was improved by the inclusion of the {\tt{reflect}} component, the resulting fit statistic ($\chi_{\rm red}^2=1.63$ with 818 dof) was still relatively poor.

Large positive residuals below about 1 keV prompted us to include a partial covering absorption component ({\tt{pcfabs}} in XSPEC). {\tt{Pcfabs}} is typically included in spectral fitting of IPs \citep{Mukai2017}. Additionally, positive residuals around 6--7 keV indicate the presence of an Fe line. A Gaussian was added to the model and freely fit to $E_{line}=6.45\pm0.04$ keV and $\sigma_{line}=0.38\pm0.06$ keV. This resulted in a $\chi_{\rm red}^2$ of 0.95 with 813 dof. Our best-fit model in XSPEC was therefore {\tt{constant*pcfabs*tbabs*(gaussian+reflect*bremss)}}, the best-fit parameters of which are listed in Table 2. However, the best-fit $\sigma_{line}$ is extraneously large and thus likely includes more than one iron line. Since the $E_{line}$ parameter is fitting to a value larger than 6.4 keV (the energy for a neutral iron fluorescence line), it is likely including the 6.7 keV (He-like) and 6.97 keV (H-like) lines as well. Therefore, we fit the spectrum to the same overall model, but with three separate Gaussian components. The line energy of each Gaussian was frozen to either 6.4, 6.7, or 6.97 keV, and the $\sigma_{line}$ for each component was frozen to 50 eV, as was done in \cite{Coughenour2022}. The best-fit parameters for this model are listed in Table \ref{bremstab}. The abundance parameter in the {\tt{reflect}} model is the abundance of elements heavier than He relative to abundances defined in \cite{Wilms2000}. The iron abundance parameter is defined in the same way and was set equal to the general abundance parameter. We froze $rel_{refl}$ to 1.0 in all cases since leaving it as a free parameter resulted in values too high to be physical. The bremsstrahlung temperatures for the single and triple Gaussian fits were not significantly different at $100_{-34}^{+97}$ and $94.4_{-39}^{+29}$ keV, respectively. While the triple Gaussian fit better constrained the bremsstrahlung temperature, it is still high for an IP.

In both bremsstrahlung fits, the best-fit abundance is lower than expected for IPs. To investigate this, we refit the data with a single Gaussian and the abundance frozen to $A=0.5$, more in line with other IPs detected by \integral\ \citep{Coughenour2022}. The resulting fit was of similar quality to the previous fits with $\chi_{\rm red}^2=0.96$ and 814 dof. The only parameters whose $1\sigma$ confidence intervals did not overlap with those of the original fit were the partial covering fraction ($f=0.68\pm0.01$) and the bremsstrahlung normalization ($N_{\rm bremss}=1.1\pm0.1\times10^{-3}$). Since our results are not affected by these low (and possibly unphysical) abundances, we continue to report our best-fit results hereafter.

Next, we replace the bremsstrahlung model with the ``post-shock region" (PSR) model or ``ipolar" model from \cite{Suleimanov2016} in order to calculate the WD mass. In addition to the WD mass, this model depends on the magnetospheric radius divided by the radius of the WD, $R_m/R_{\rm WD}$. The magnetospheric radius is the point at which the accretion disc is truncated by the WD's magnetic field. By assuming that $R_m$ is equal to the co-rotation radius of the WD, we can set $R_m=(\frac{GMP^2}{4\pi^2})^{1/3}$ where $P$ is the spin period of the WD, which we set to 304.4 s \citep{Suleimanov2016}. $R_{\rm WD}$ can then be calculated using the well-known WD equation of state from \cite{Nauenberg1972} relating WD mass and radius. The $R_m$ parameter in the ipolar model can therefore be linked to the mass parameter in XSPEC according to these two equations, as was done in \cite{Tomsick2023} and \cite{Coughenour2022}.

\begin{figure*}
\begin{center}
    \includegraphics[scale=.5]{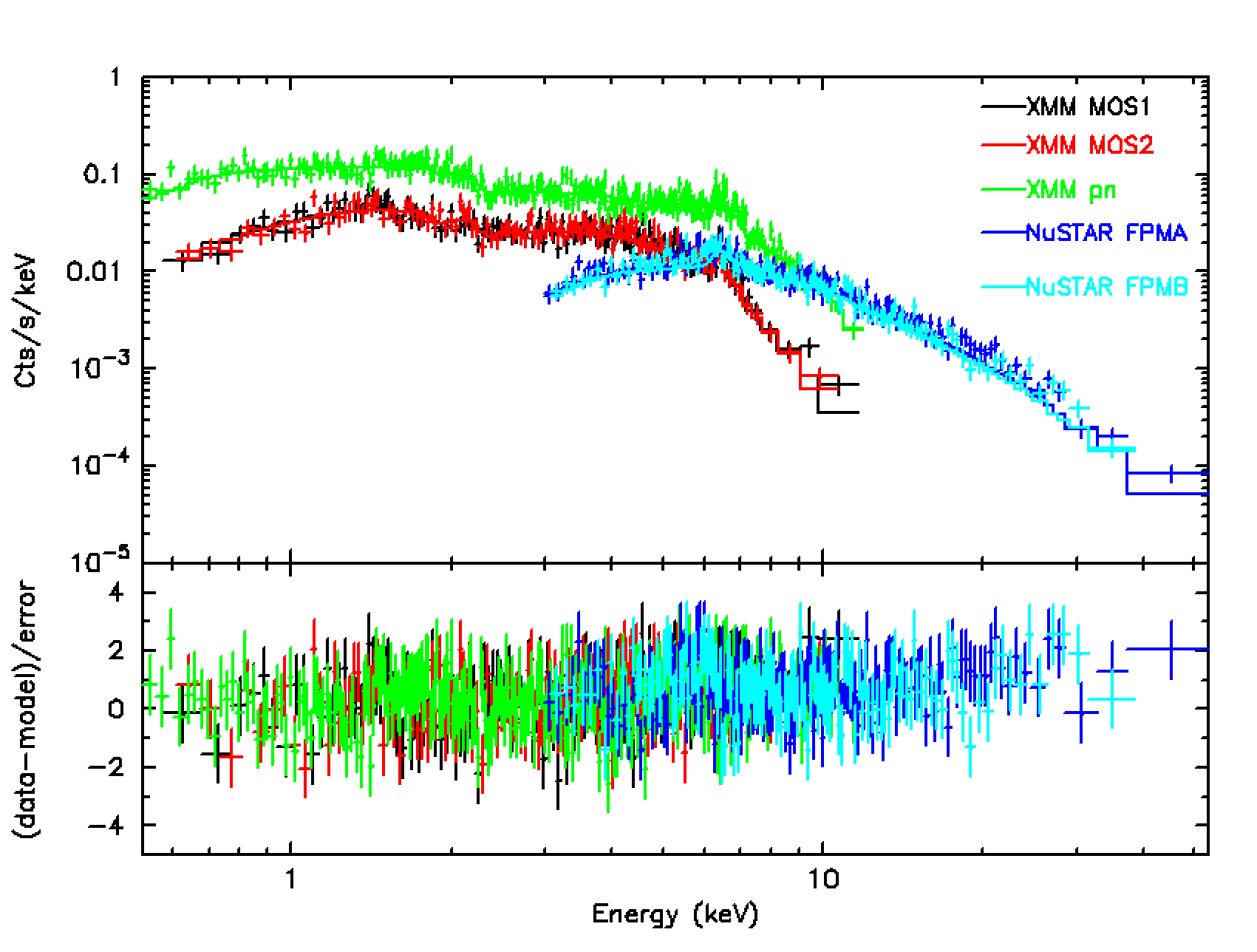}
    \caption{\xmm\ and \nustar\ joint spectrum (with residuals) fit to the PSR model and three narrow Gaussians. The best-fit parameters for this model are listed in Table \ref{psrtab}}
    \label{spec}
\end{center}
\end{figure*}

The best-fit parameters of the ipolar fit with three Gaussians are displayed in Table \ref{psrtab}. The spectrum and corresponding residuals are shown in Figure \ref{spec}. The best-fit mass was 1.4 \ms, the Chandrasekhar limit and the maximum value allowed in XSPEC. This gives $R_m=50.2$ $R_{\rm WD}$. The $1\sigma$ lower limit on the mass was 1.36 \ms. The unabsorbed flux was calculated by applying the {\tt cflux} convolution model to the additive model components. The full model was therefore {\tt {constant * pcfabs * tbabs * cflux * (gaussian + gaussian + gaussian + reflect * atable\{ipolar.fits\})}}. The energy range for the {\tt cflux} component was set to 0.5 -- 12 keV for the \xmm\ spectra and 3 -- 79 keV for the \nustar\ spectra. Since the cross-normalization constant factors all fit to $\sim$1, the best-fit flux is applicable across instruments. The 0.5 -- 12 keV flux is $6.36_{-0.07}^{+0.06}\times10^{-12}$ \fluxcgs. The 3 -- 79 keV flux is $1.43_{-0.02}^{+0.01}\times10^{-11}$ \fluxcgs. Assuming a source distance of 3 kpc \citep{BailerJones2021}, these fluxes give luminosities of $6.85\pm0.07\times10^{33}$ \lumcgs\ and $1.54_{-0.02}^{+0.01}\times10^{34}$ \lumcgs\ in the 0.5 -- 12 keV and 3 -- 79 keV bands, respectively. The flux in the 17 -- 60 keV band is $7.0\pm0.1\times10^{-12}$ \fluxcgs\ as compared to $5.2\pm0.8\times10^{-12}$ \fluxcgs\ reported in \cite{Tomsick2021} as measured by \integral.

Again, the best-fit abundance for the ipolar fit with three Gaussians is smaller than expected. We try the same procedure as before, refitting the data with abundance frozen to $A=0.5$. This fit gave a $\chi_{\rm red}^2=0.98$ and 814 dof. The parameters whose $1\sigma$ confidence intervals did not overlap with those displayed in Table \ref{psrtab} were $N_{\rm H,pc}=6.2\pm0.5\times10^{22}$ cm$^{-2}$, the partial covering fraction $=0.70\pm0.01$, and $\cos i =0.3\pm0.1$. Although the 1$\sigma$ upper limit on the mass reached 1.4 \ms, the mass was best fit to 1.36 \ms with a lower bound of 1.33 \ms.

\section{Discussion}
\label{discussion}
The classification of \src\ determined by \cite{Tomsick2021} is supported by the above spectral analysis due to its spectra being well fit to the PSR model. Furthermore, the detection of a strong iron complex in the 6--7 keV range is consistent with the X-ray properties of IPs, as is the detection of a spin period in the optical and X-ray bands.

Based on time-series photometry conducted by \cite{Halpern2022}, a signal in the optical power spectrum of \src\ was found at $152.49\pm0.02$~s. Through the X-ray timing analysis described in Section 3, we find a period of $P=304.4\pm0.3$ s, nearly twice the previously reported signal. We therefore conclude that $P=304.4\pm0.3$~s is the true WD spin period, and that the $152.49\pm0.02$~s signal is a harmonic of the fundamental period. It is fairly common for the initially discovered period to turn out to be a harmonic rather than the fundamental frequency. For example, in the case of the IP V2306 Cygni (WGA J1958.2+3232), observations from both ASCA and the Astrophysical Observatory of Catania (OACT) observed a period of about 733 s \citep{Israel1998, Uslenghi2000}. However, OACT also observed a peak in the periodograms at 1466 s, twice the previously noted period.

\cite{Norton1999} describes how two-pole accretion does not exclusively lead to a double-peaked pulse profile. Instead, two-pole accretion may lead to either a single or double-peaked pulse profile, depending on the strength of the white dwarf's magnetic field. If the magnetic field is weak, material travels along the field lines beginning closer to the surface of the WD. This produces a larger accretion region, and thus the optical depth across the accretion column can be greater than the optical depth up the accretion column. Therefore, a double peaked profile will be produced since there will be a maximum in received flux at the two observing points that align with the magnetic field lines. This double peaked effect, according to \cite{Norton1999}, is predominantly seen in IPs with a WD spin period less than $\sim$700 s, as is the case for \src. The idea that the accretion disc is truncated close to the WD surface seems to contradict our finding that $R_m=50.2$ $R_{\rm WD}$. However, given how massive \src\ is, the WD is extremely compact leading to a higher $R_m/R_{\rm WD}$. It may therefore be the case that the absolute value of $R_m$ is more significant to the pulse profile shape than the ratio of $R_m$ to $R_{\rm WD}$. While a double peak is observed in the 0.5 -- 2 keV \xmm\ pulse profile, only a single peak is observed in the 2 -- 10 keV band. This difference in pulse profiles may be due to the fact that, although the geometric effects are similar at both energies, the photoelectric absorption varies.

We can use the best-fit $R_m$ to estimate $\dot{M}$, the mass accretion rate and $B$, the surface magnetic field at the accretion region. We use the formula, $L=GM\dot{M}(\frac{1}{R}-\frac{1}{R_m})$ and $L=1.45\times10^{34}$ \lumcgs to find $\dot{M}=1.3\times10^{16}$ g/s. We then use this value to calculate $B$ using equation 8 in \cite{suleimanov19} which assumes $R_m$ is proportional to the Alfv\'en radius ($R_m=\Lambda R_A$). Assuming $\Lambda=0.5$ (as is done in \cite{suleimanov19}), we find $B=94$ MG. If we use the lower bound value on the best-fit mass ($M=1.36 M_{\odot}$), we find $R_m=34.9$ $R_{\rm WD}$ and $B=39$ MG. We also use equation 3 in \cite{Norton&Watson1989} to calculate the magnetic moment, $\mu=1.6\times10^{32}$ G cm$^3$. Here, we have assumed $\phi=1$, or that the magnetosphere is spherically symmetric. 

The estimated magnetic moment of \src, $\mu=1.6\times10^{32}$ G cm$^3$, given $M=1.36$ $M_{\odot}$, is consistent with the range of expected values ($10^{32-33}$ G cm$^3$) found by \cite{Norton2004} for most IPs. Polars, in contrast, generally have higher WD magnetic moments ($\mu \goe 10^{33}$ G cm$^3$). However, \cite{Norton2004} suggests that low magnetic moment systems (such as \src) may still evolve to polars given a long orbital period ($P_{\rm orb}>3$ hr). Follow-up observations confirming the orbital period of \src\ could therefore be useful in determining its evolution.

It is possible that the 3 hour peak observed in the \nustar\ and \xmm\ light curves is the orbital period. The large modulation due to this potential orbital period is evident in the \xmm\ light curves shown in Figure \ref{pn_05-2_2_10LC}. However, the length of the observation only spans $\sim$2.1 times this period, so this candidate period should be verified through additional observations. If this is the true orbital period though, it would be comparable to that of the IP BG CMi which has an orbital period of $\sim$3.25 hr \citep{deMartino1995}. \cite{Parker2005} suggests that orbital modulation in IPs is due to photoelectric absorption at the accretion disc. \cite{Parker2005} predicts that the inclination angle required for orbital modulation to be seen is greater than $60^{\circ}$. According to our spectral modeling, the angle between our line of sight and the accretion column is $\cos i> 0.60$ or $i\simlt50^{\circ}$. There may therefore be a misalignment between the accretion column and the normal to the orbital plane. However, \cite{Mukai2015} discusses the degeneracy between the angle of the accretion column and the reflection amplitude, which makes it difficult to determine either parameter through spectral analysis alone. 

The continued detection of high mass WDs by \integral\ such as IGR J14091-6108, IGR J18007-4146, and IGR J15038-6021 (\cite{Tomsick2016, Coughenour2022, Tomsick2023}) may indicate that WDs in CVs gain mass throughout accretion-nova cycles. Simulations of high mass WDs ($M > 0.6$ \ms) by \cite{Starrfield2020} found that these WDs can accrete more mass than they lose via classical novae events. Once a WD gains enough mass to reach the Chandrasekhar limit ($\sim$1.4 \ms), they may lead to Type Ia supernovae, significant for their role in cosmology as standard candles. Another possible solution to the WD mass problem is that low mass WDs in CVs lose angular momentum and merge with their donor stars, leaving a greater number of CVs with higher mass WDs to be observed \citep{Schreiber2016}.

\begin{figure*}
\begin{center}
    \includegraphics[width=15cm]{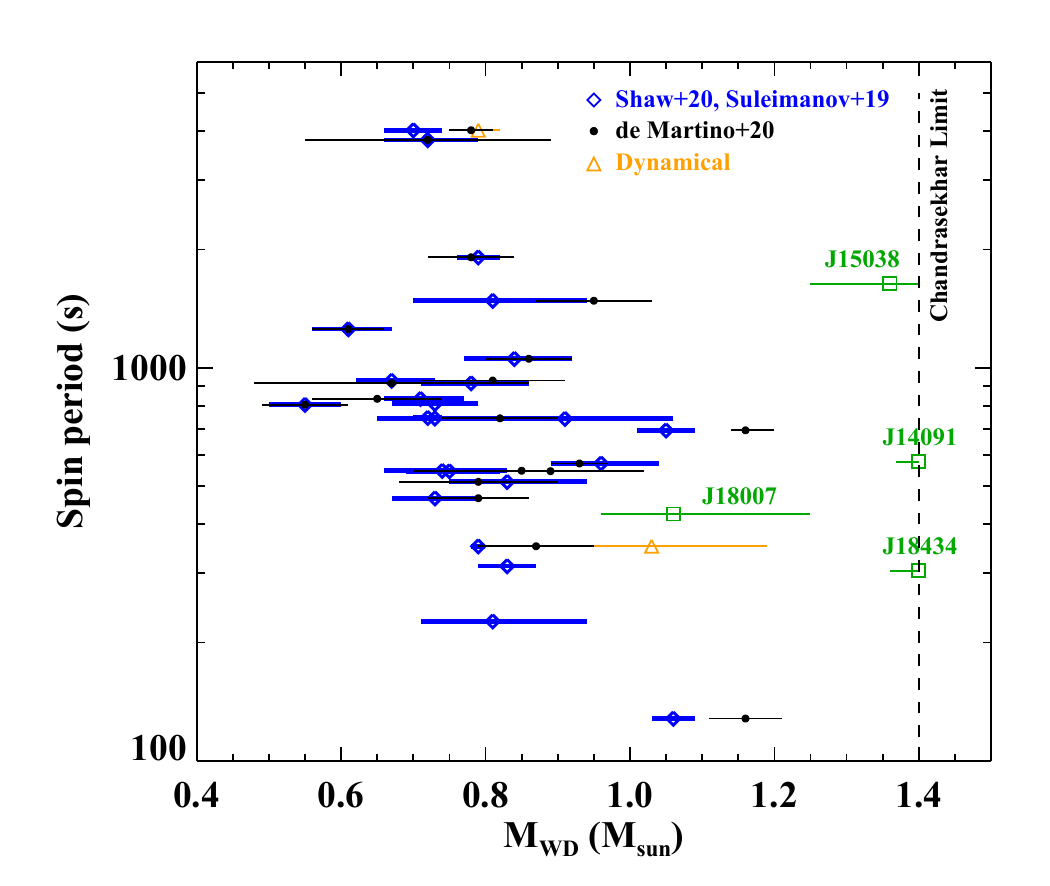}
    \caption{Plot of WD masses and spin periods of IPs. This graph is Figure 10 from \protect\cite{Tomsick2023} with data from \src\ included. The IPs marked in green are from our studies of IGR sources: J14091 \protect\citep{Tomsick2016}, J18007 \protect\citep{Coughenour2022}, J15038 \protect\citep{Tomsick2023}, and J18434 (this work).}
    \label{pspinandmass}
\end{center}
\end{figure*}

Due to its high energy band, \integral\ is exceptionally well equipped to search for high mass IPs. Figure \ref{pspinandmass} shows the WD masses and spin periods of various IPs as measured by \cite{shaw20}, \cite{suleimanov19}, \cite{rk11}, \cite{demartino20}, and our studies of IGR sources. \integral's ability to detect high mass IPs is illustrated in Figure \ref{pspinandmass} where the three IPs with masses closest to the Chandrasekhar limit are those detected by \integral. 

\section*{Acknowledgements}
JH acknowledges support from NASA under award number 80GSFC21M0002. MC acknowledges financial support from the Centre National d’Etudes Spatiales (CNES). JAT and AJ acknowledge partial support from NASA under award numbers 80NSSC21K0064 and 80NSSC22K0055.
\section*{Data Availability}
Data used in this paper are available through NASA’s HEASARC.
\label{lastpage}
\bibliography{main.bib}
\bibliographystyle{aasjournal}
\clearpage

\begin{table*}
\caption{Details of Observations of \src}
\begin{minipage}{\linewidth}
\begin{tabular}{cccccc} \hline \hline
Observatory & ObsID & Instrument & Start Time (UT) & End Time (UT) & Exposure (ks) \\ \hline
\xmm & 0890620201 & MOS1 & 2022 March 12, 16:57:00 & 2022 March 13, 00:20:00 & 25.7 \\
" & " & MOS2 & " & " & 26.0 \\ 
" & " & pn & " & " & 18.3 \\
\nustar & 30760002002 & FPMA & 2022 March 12, 15:16:09 & 2022 March 13, 15:16:09 & 40.2 \\
" & " & FPMB & " & " & 39.8 \\
\hline
\end{tabular}
\end{minipage}
\label{obstab}
\end{table*}
\clearpage
\clearpage
\clearpage
\clearpage

\clearpage
\clearpage

\begin{table*}
\caption{Spectral results for Bremsstrahlung model fits}
\begin{minipage}{\linewidth}
\begin{tabular}{cccc} \hline \hline
Parameter\footnote{The errors on the parameters are $1\sigma$ confidence intervals.} &
Units &
1 Gaussian\footnote{The full XSPEC model is {\tt{constant*pcfabs*tbabs*(gaussian+reflect*bremss)}}} &
3 Gaussians\footnote{The full XSPEC model is {\tt{constant*pcfabs*tbabs*(gaussian+gaussian+gaussian+reflect*bremss)}}} \\ \hline
$N_{\rm H}$ & $10^{22}$ cm$^{-2}$ & $0.17_{-0.02}^{+0.05}$ & $0.17_{-0.02}^{+0.06}$ \\
$N_{\rm H,pc}$ & $10^{22}$ cm$^{-2}$  & $4.7_{-0.4}^{+1.0}$ & $5.46_{-1.2}^{+0.95}$ \\
pc fraction & --- & $0.52_{-0.01}^{+0.03}$ & $0.53\pm0.04$ \\
\hline
$E_{\rm line1}$ & keV  & $6.45\pm0.04$ & $6.4$\footnote{Frozen} \\
$\sigma_{\rm line1}$ & keV  & $0.38\pm0.06$ & $0.05^d$ \\
$N_{\rm line1}$ & ph cm$^{-2}$ s$^{-1}$  & $(2.66_{-0.29}^{+0.23})\times 10^{-5}$ & $(1.24\pm0.1)\times10^{-5}$ \\
$EW_{\rm line1}$ & eV & $505_{-55}^{+43}$ & $212\pm17$ \\
\hline
$E_{\rm line2}$ & keV  & --- & $6.7^d$ \\
$\sigma_{\rm line2}$ & keV & --- & $0.05^d$ \\
$N_{\rm line2}$ & ph cm$^{-2}$ s$^{-1}$  & --- & $(0.4\pm0.1)\times10^{-5}$ \\
$EW_{\rm line2}$ & eV & --- & $53_{-13}^{+14}$ \\
\hline
$E_{\rm line3}$ & keV  & --- & $6.97^d$ \\
$\sigma_{\rm line3}$ & keV  & --- & $0.05^d$ \\
$N_{\rm line3}$ & ph cm$^{-2}$ s$^{-1}$  & --- & $(0.4\pm0.1)\times10^{-5}$ \\
$EW_{\rm line3}$ & eV  & --- & $72\pm18$ \\
\hline
$rel_{\rm refl}$ & ---  & 1.0$^d$ & 1.0$^d$ \\
$A$ & --- & $0.007_{-0.003}^{+0.09}$ & $0.006_{-0.003}^{+0.15}$ \\
$A_{\rm Fe}$\footnote{Tied to parameter $A$} & --- & $0.007$ & $0.006$ \\
$\cos{i}$ & --- &  $0.75_{-0.5}^{+0.1}$ & $>0.28$ \\ 
\hline
$kT$ & keV & $100.0_{-34}^{+97}$ & $94.4_{-39}^{+29}$\\
$N_{\rm bremss}$ & --- & $(6.4^{+0.5}_{-0.2})\times 10^{-4}$ & $(6.3_{-0.6}^{+0.4})\times 10^{-4}$\\
\hline
C$_{\rm MOS1}$ & --- & 1.0$^d$ & 1.0$^d$ \\
C$_{\rm MOS2}$ & --- &  $0.97\pm0.02$ & $0.97\pm0.02$ \\ 
C$_{\rm pn}$ & --- &  $0.99\pm0.02$ & $0.99\pm0.02$ \\
C$_{\rm FPMA}$ & --- &  $1.19_{-0.02}^{+0.03}$ & $1.18\pm0.03$ \\
C$_{\rm FPMB}$ & --- & $1.20\pm0.03$ & $1.19\pm0.03$ \\
\hline
$\chi_{\rm \nu}^{2}/$(dof) & --- &  0.95(813) & 0.96(813)\\
\hline
\end{tabular}
\end{minipage}
\label{bremstab}
\end{table*}

\begin{table*}
\caption{Spectral results for PSR model fit}
\begin{minipage}{\linewidth}
\begin{tabular}{ccc} \hline \hline
Parameter\footnote{The errors on the parameters are $1\sigma$ confidence intervals.} &
Units &
PSR Model\footnote{The full XSPEC model is {\tt{constant*pcfabs*tbabs*cflux*(gaussian+gaussian+gaussian+reflect*atable\{ipolar.fits\})}}} \\ \hline
$N_{\rm H}$ & $10^{22}$ cm$^{-2}$ & $0.20_{-0.02}^{+0.03}$  \\
$N_{\rm H,pc}$ & $10^{22}$ cm$^{-2}$  & $4.46_{-0.7}^{+0.9}$ \\
pc fraction & --- & $0.56_{-0.03}^{+0.04}$  \\
\hline
$E_{\rm line1}$ & keV & $6.4$\footnote{Frozen} \\
$\sigma_{\rm line1}$ & keV  & $0.05^c$ \\
$N_{\rm line1}$ & ph cm$^{-2}$ s$^{-1}$  & $(1.2\pm0.1)\times 10^{-5}$  \\
$EW_{\rm line1}$ & eV & $204\pm17$ \\
\hline
$E_{\rm line2}$ & keV & $6.7^c$ \\
$\sigma_{\rm line2}$ & keV & $0.05^c$ \\
$N_{\rm line2}$ & ph cm$^{-2}$ s$^{-1}$  & $(0.38\pm0.1)\times10^{-5}$ \\
$EW_{\rm line2}$ & eV  & $51_{-13}^{+14}$ \\
\hline
$E_{\rm line3}$ & keV  & $6.97^c$ \\
$\sigma_{\rm line3}$ & keV   & $0.05^c$ \\
$N_{\rm line3}$ & ph cm$^{-2}$ s$^{-1}$   & $(0.36\pm0.1)\times10^{-5}$ \\
$EW_{\rm line3}$ & eV & $65_{-18}^{+19}$ \\
\hline
$rel_{\rm refl}$ & ---  & 1.0$^c$ \\
$A$ & --- & $0.02\pm0.01$  \\
$A_{\rm Fe}$\footnote{Tied to parameter $A$} & --- & $0.02$  \\
$\cos{i}$ & --- &  $>0.60$  \\ 
\hline
$M_{\rm WD}$ & \ms & $>1.36$\\
$R_{\rm m}$\footnote{Linked to $M_{\rm WD}$ via equations 3 and 4 in \cite{Suleimanov2016}} & $R_{\rm WD}$ & 50.2\\
$N_{\rm PSR}$ & --- & $(2.88^{+5}_{-0.2})\times 10^{-30}$\\
\hline
C$_{\rm MOS1}$ & --- & 1.0$^c$ \\
C$_{\rm MOS2}$ & --- &  $0.97\pm0.02$ \\ 
C$_{\rm pn}$ & --- &  $0.99\pm0.02$  \\
C$_{\rm FPMA}$ & --- &  $1.18\pm0.03$ \\
C$_{\rm FPMB}$ & --- & $1.19\pm0.03$  \\
\hline
$\chi_{\rm \nu}^{2}/$(dof) & --- &  0.96(813)\\
\hline
\end{tabular}
\end{minipage}
\label{psrtab}
\end{table*}

\clearpage

\clearpage

\end{document}